\documentclass{article}
\usepackage{ijcai26}

\usepackage{times}
\usepackage{soul}
\usepackage{url}
\usepackage[hidelinks]{hyperref}
\usepackage[utf8]{inputenc}
\usepackage[small]{caption}
\usepackage{graphicx}
\usepackage{amsmath}
\usepackage{amsthm}
\usepackage{booktabs}
\usepackage{algorithm}
\usepackage{algorithmic}
\usepackage[switch]{lineno}

\usepackage{mathtools}

\usepackage{siunitx}
\usepackage{booktabs}
\usepackage{multirow}
\usepackage{amssymb}    
\usepackage{bm}         

\usepackage{makecell}

\title{Detect, Attend and Extract: Keyword Guided Target Speaker Extraction}

\author{
Haoyu Li$^{1,2,   
 *}$
\and
Yu Xi$^{2,4,*}$\and
Yidi Jiang$^{3}$\and
Shuai Wang$^{1,6,\dagger}$\and 
Kate Knill$^{4}$\and
Mark Gales$^{4}$\and \\
Haizhou Li$^{5,6}$\And
Kai Yu$^{2,\dagger}$
\affiliations
$^1$School of Intelligence Science and Technology, Nanjing University, Suzhou, China \\
$^2$X-LANCE Lab, MoE Key Lab of Artificial Intelligence, School of Computer Science,\\Shanghai Jiao Tong University, Shanghai, China \\
$^3$National University of Singapore, Singapore \\
$^4$ALTA Institute, Machine Intelligence Lab, Department of Engineering, University of Cambridge, UK\\
$^5$The Chinese University of Hong Kong, Shenzhen, China \\
$^6$Shenzhen Loop Area Institute, Shenzhen, China
\emails
\{haoyu.li.cs,yuxi.cs\}@sjtu.edu.cn,
shuaiwang@nju.edu.cn,
kai.yu@sjtu.edu.cn
}

\begin{document}

\maketitle

\def\thefootnote{$\dagger$}\footnotetext{~indicates the corresponding authors.}\def\thefootnote{\arabic{footnote}}

\begin{abstract}
    Target speaker extraction (TSE) aims to extract the speech of a target speaker from mixtures containing multiple competing speakers. Conventional TSE systems predominantly rely on speaker cues, such as pre-enrolled speech, to identify and isolate the target speaker. However, in many practical scenarios, clean enrollment utterances are unavailable, limiting the applicability of existing approaches.
    In this work, we propose DAE-TSE, a keyword-guided TSE framework that specifies the target speaker through distinct keywords they utter. By leveraging keywords (i.e., partial transcriptions) as cues, our approach provides a flexible and practical alternative to enrollment-based TSE. DAE-TSE follows the Detect-Attend-Extract (DAE) paradigm: it first detects the presence of the given keywords, then attends to the corresponding speaker based on the keyword content, and finally extracts the target speech.
    Experimental results demonstrate that DAE-TSE outperforms standard TSE systems that rely on clean enrollment speech.
    To the best of our knowledge, this is the first study to utilize partial transcription as a cue for specifying the target speaker in TSE, offering a flexible and practical solution for real-world scenarios. Our code\footnote{\url{https://github.com/GnafiY/DAE-TSE}} and demo page\footnote{\url{https://gnafiy.github.io/DAE-TSE_demo}} are now publicly available.
\end{abstract}

\section{Introduction}
Target speaker extraction (TSE) aims to isolate the speech of the target speaker from a mixture of multiple speakers, also known as the cocktail party problem~\cite{nntse}. In doing so, we need to find ways to inform the system who the target speaker is by providing a reference cue. There have been studies on different reference cues, such as pre-enrolled reference speech~\cite{icassp2023-yanglei-ultra_short_ref,icassp2025-zhangke-multi_level_speaker_emb,icassp2024-pengjunyi-ssl_tse,is2024-liuyun-cl_ref_speech_tse,is2024-zhangyiru-or_tse,is2024-heo-centroid_speaker_emb}, target-speaker lip movements~\cite{icassp2023-linjiuxin-av_sepformer,arxiv-taoruijie-lip_tse,is2024-wutianci-lip_tse,icassp2024-lijunjie-lip_tse}, acoustic spatial information~\cite{icassp2024-heshulin-3s_tse,icassp2024-wangyichi-spatial_tse,icassp2024-suntianchi-directional_tse,is2024-ashutosh-nldse}, and language information~\cite{asru2021-borsdorf-languagetse,icassp2022-borsdorf-languagetse,apsipa_asc_2025-sinan-languagetse}. 

\begin{figure}[t!]
    \centering
      \includegraphics[scale=0.42]{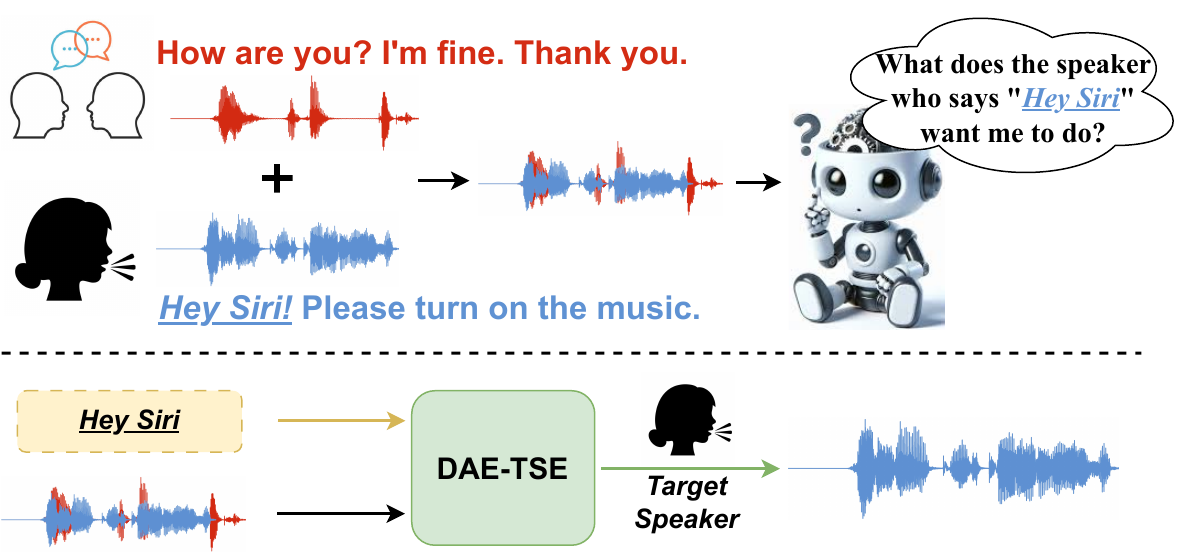}
      \caption{Illustration of the application scenario and objectives of the proposed DAE-TSE framework. In multi-talker scenarios, DAE-TSE aims to extract the speech of the target speaker who uttered the given keywords, such as ``Hey Siri," from the mixture.}
    \label{fig:illustration}
\end{figure}

However, in dynamic environments such as ad-hoc meetings or voice assistant interactions, obtaining pre-registered voiceprints for all potential speakers is often unfeasible. Furthermore, relying on sustained auxiliary cues poses significant challenges; it is impractical to require a target speaker to maintain a fixed spatial location for reliable Direction of Arrival (DOA) estimation or to ensure that their lip movements remain constantly visible to a camera. In such unconstrained contexts, text-based enrollment offers a compelling solution.
Prior research has explored extraction using \textit{natural language descriptions}, known as \textit{Language-queried Audio Source Separation (LASS)}~\cite{is2022-liuxubo-liuxubo,iclr2023-donghaowen-clipsep,xianghao-typing,taslp2024-mahao-clapsep,arxiv2025-meta-sam}. These systems isolate sources based on descriptive attributes (e.g., ``female speaker'', ``the loudest speaker'').
Conversely, \textit{keyword-guided extraction}, which uses \textit{partial transcription} as a reference, remains underexplored despite its immense potential. As illustrated in Figure~\ref{fig:illustration}, a user may wish to retrieve speech from a participant based on salient keywords they mentioned (e.g., specific topics). This framework offers greater flexibility by enabling full speech extraction using only partial transcription cues, thereby eliminating the need for cumbersome pre-enrollment procedures.
It is crucial to distinguish between natural language descriptions and keyword cues. LASS systems use descriptive prompts as semantic attributes. In contrast, keyword-guided TSE must identify the target speaker by aligning the acoustic content with the specific words they speak. Existing instruction-following systems, such as LLM-TSE~\cite{xianghao-typing}, treat text as a semantic instruction rather than explicitly performing keyword-guided alignment and identification.

The primary challenge in keyword-guided TSE lies in leveraging local linguistic information (keywords) to infer a global speaker representation. Unlike traditional TSE, which derives global embeddings from enrolled speech, our method must bridge content-speaker alignment to generate speaker-aware representations.
To address this, we investigate the design of a partial keyword-guided TSE system that incorporates an effective cue encoder to derive target speaker embeddings conditioned on the provided transcription. Specifically, we propose the Detect-Attend-Extract (DAE) keyword-guided TSE framework, which consists of two main components: (1) a Keyword-guided Cue Encoder (KCE) that generates target speaker embeddings from the speech mixture and the keyword transcription, and (2) a TSE backbone that extracts the complete target speech. The KCE is trained via a joint Automatic Speech Recognition (ASR) and Speaker Verification (SV) objective, employing a cross-attention mechanism to align the mixture with keywords. Such a design enables DAE-TSE to determine the presence of keywords and localize their timestamps in the mixture when they exist.

Overall, DAE-TSE first detects the presence of the keywords in the mixture, then attends to the target speaker based on the keyword context, and finally extracts the target speech using the TSE backbone.
Our core contributions are:
\begin{itemize}
    \item We propose the Detect-Attend-Extract TSE (DAE-TSE) framework, which extracts target-speaker speech using only short keyword transcripts, eliminating the need for pre-enrolled speech.
    \item We propose a Keyword-guided Cue Encoder (KCE) under joint ASR-SV training, aligning textual and acoustic features via cross-attention to enable keyword detection, localization, and global speaker embedding extraction.
    \item Experimental results demonstrate that DAE-TSE surpasses competitive baselines while utilizing only 28.4\% of the complete transcription, attaining a keyword localization error of $\sim$100~ms.
\end{itemize}

\section{DAE-TSE: Keyword Guided Target Speaker Extraction}
\subsection{Target Speaker Extraction Fundamentals}

Consider a mixture $\mathbf{x} \in \mathbb{R}^{L}$ consisting of $L$ samples of the target speaker's waveform $\mathbf{y} \in \mathbb{R}^{L}$ and $N$ interfering sources $\mathbf{n}_i \in \mathbb{R}^{L}$, i.e.,
\begin{equation}
\mathbf{x}= \mathbf{y}+ \sum\nolimits_{i=1}^{N}\mathbf{n}_i,
\end{equation}
TSE aims to recover $\mathbf{y}$ from $\mathbf{x}$.
Target speaker extraction relies on a conditioning cue $\mathbf{q}$ that encodes speaker identity or content, typically formulated as a fixed vector or temporal sequence. Mainstreaming TSE systems derive an embedding $\mathbf{q}$ from a clean enrollment utterance via a pre-trained speaker encoder. Alternatively, categorical cues such as spatial layout (near vs. far-field) or gender (male vs. female) are presented as learnable embeddings. In audio-visual TSE, $\mathbf{q}$ is a latent stream extracted from lip motion or facial features.

\begin{figure*}[!t]
    \centering
      \includegraphics[scale=0.54]{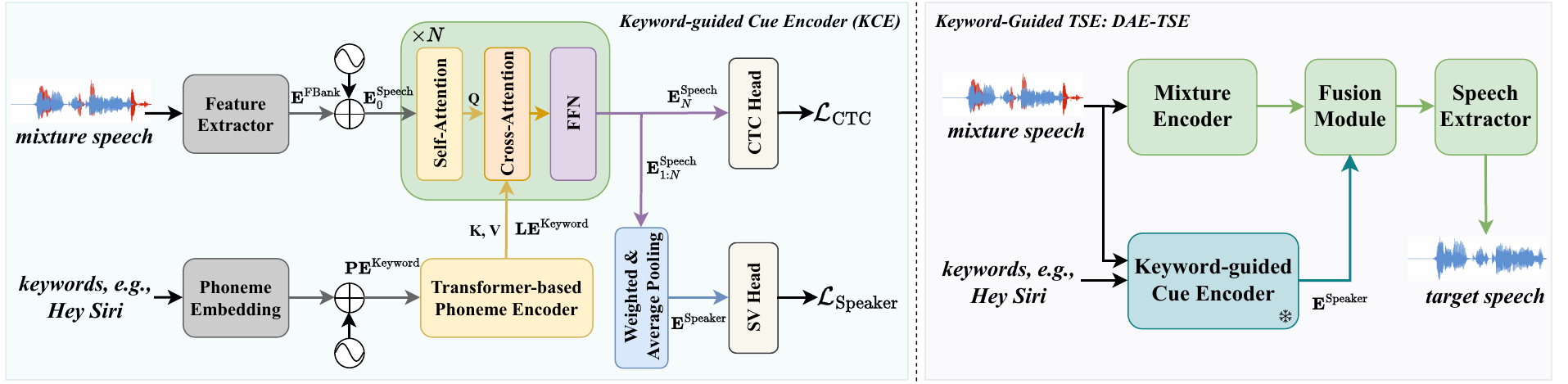}
      \caption{Overview of the proposed DAE-TSE framework. (1) The left panel depicts the architecture of the Keyword-guided Cue Encoder (KCE), which is jointly optimized via ASR and SV losses and processes the mixture speech and keywords to yield target-speaker cue embeddings. (2) The right panel shows the DAE-TSE training pipeline leveraging the pretrained KCE, which extracts the target speech from mixture inputs guided by cue embedding $\mathbf{E}^{\text{Speaker}}$.}
    \label{fig:overview}
\end{figure*}

A standard TSE system typically consists of two major components: a cue encoder and a speech extraction module. The latter further contains a mixture encoder, a fusion module, and a target decoder~\cite{nntse}. In general, the TSE model extracts the target speech $\hat{\mathbf{y}}$ from the mixture $\mathbf{x}$ conditioned on the cue $\mathbf{q}$:
\begin{equation}
    \hat{\mathbf{y}} = \mathcal{F}_{\boldsymbol{\theta}} \left( \mathbf{x}, \mathbf{q} \right),
\end{equation}
where $\mathcal{F}_{\boldsymbol{\theta}}(\cdot)$ denotes the TSE model parameterized by $\boldsymbol{\theta}$.
The right panel of Figure~\ref{fig:overview} depicts the overall architecture of DAE-TSE, comprising a keyword-guided cue encoder and a speech extraction backbone.

\subsection{Keyword-guided Cue Encoder for DAE-TSE}
\subsubsection{Architecture}
\label{sec:kce_module}

In this section, we introduce the Keyword-guided Cue Encoder (KCE) for DAE-TSE, which leverages a few keywords to extract target speaker embeddings from the mixture. 
The KCE adopts a Transformer-based architecture that takes keywords and the speech mixture as dual inputs. Through cross-attention mechanisms, it aligns textual and acoustic representations to produce a compact cue embedding for TSE.

As illustrated in Figure~\ref{fig:overview}, the input keywords are first converted into phoneme-based embeddings $\mathbf{PE}^\text{Keyword} \in \mathbb{R}^{L_\text{kw} \times D_\text{kw}}$, where $L_\text{kw}$ denotes the length of the phoneme sequence, and $D_\text{kw}$ is the phoneme embedding dimension. Keyword embeddings are then passed through a Transformer-based encoder to obtain the latent representations $\mathbf{LE}^\text{Keyword} \in \mathbb{R}^{L_\text{kw} \times D}$, where $D$ is the output dimension of the Transformer. The overall operation is formulated as:
\begin{equation}
\mathbf{LE}^\text{Keyword} = \text{Transformers}(\mathbf{PE}^\text{Keyword}).
\end{equation}
For the speech input, the mixture is first transformed into log-Mel filterbank (FBank) features, denoted as $\mathbf{E}^\text{FBank} \in \mathbb{R}^{T \times D_\text{fb}}$, where $T$ is the number of frames and $D_\text{fb}$ is the acoustic feature dimension. These features are combined with positional embeddings to form the initial input to the Transformer-based speech encoder, represented as $\mathbf{E}^\text{Speech}_{0}$. To incorporate keyword information, each Transformer block is augmented with a cross-attention module that fuses acoustic and textual representations. Specifically, in the $i$-th Transformer block, the speech features serve as the queries:
\begin{equation}
\mathbf{Q}^\text{Speech}_i = \text{Attention}(\mathbf{E}^\text{Speech}_{i-1}, \mathbf{E}^\text{Speech}_{i-1}, \mathbf{E}^\text{Speech}_{i-1}),
\end{equation}
and the output of the block is computed as:
\begin{equation}
\resizebox{0.91\columnwidth}{!}{
$\displaystyle \mathbf{E}^\text{Speech}_{i}, \mathbf{M}_{i} = \text{FFN}(\text{Attention}(\mathbf{Q}^\text{Speech}_i, \mathbf{LE}^\text{Keyword}, \mathbf{LE}^\text{Keyword})),$}
\label{eq:attention_map}
\end{equation}
where FFN denotes a feed-forward network, $\mathbf{M}_{i}$ is the attention map, and $\mathbf{LE}^\text{Keyword}$ represents the keyword-derived key and value embeddings used for cross-attention. After processing through all $N$ layers of the speech encoder, we obtain the final output:
\begin{equation}
\mathbf{E}^\text{Speech} = [\mathbf{E}^\text{Speech}_{1}, \mathbf{E}^\text{Speech}_{2}, \cdots, \mathbf{E}^\text{Speech}_{N}] \in \mathbb{R}^{N \times T \times D}.
\end{equation}

\subsubsection{Training Criteria} 
To train KCE, we adopt a multi-task learning paradigm that combines Automatic Speech Recognition (ASR) and Speaker Verification (SV). Specifically, the ASR task predicts the transcription containing the keywords, while the SV predicts the speaker identity who utters the keyword.

Given the mixture speech $\mathbf{x}$, assuming the corresponding target transcription is $\mathbf{y}^\text{Trans}$, and the target speaker identity is $\mathbf{y}^\text{Spk}$, the Connectionist Temporal Classification (CTC)-based ASR loss can be represented as:
\begin{align}
\mathcal{L}_{\mathrm{CTC}}\left(\mathbf{x},\mathbf{y}^\text{Trans}\right)=&-\log p\left(\mathbf{y}^\text{Trans}|\mathbf{x}\right) \\ =&-\log \, \sum_{\mathclap{\hspace{0.5em} \mathbf{\pi_\mathrm{CTC}} \in \mathcal{B}_\mathrm{CTC}^{-1}(\mathbf{y}^\text{Trans})}}\hspace{0.5em}p\left(\bm{\pi}_\mathrm{CTC}|\mathbf{x}\right).
\end{align}
Here, $\mathcal{B}_{\mathrm{CTC}}$ maps valid CTC alignments $\bm{\pi}_{\mathrm{CTC}}$ (including the blank token) to the label sequence $\mathbf{y}^\text{Trans}$, while $\mathcal{B}_{\mathrm{CTC}}^{-1}$ represents its inverse mapping. In contrast to ASR, which typically uses the final layer representation $\mathbf{E}^\text{Speech}_{N}$ to calculate the CTC loss, we used a weighted layer pooling to obtain the speaker information from each layer of the speech encoder dynamically. Specifically, a trainable weight vector $\mathbf{w} = [w_{1}, w_{2}, \cdots, w_{N}] \in \mathbb{R}^{N}$ is applied to fuse the keyword-biased speech representations across Transformer layers. The fused feature is computed as:
\begin{equation}
\mathbf{E}^\text{Speech}_\text{sum} = \sum_{i=1}^{N} \mathbf{w}_{i} \mathbf{E}^\text{Speech}_{i} \in \mathbb{R}^{T\times D}.
\end{equation}
To obtain a condensed speaker representation, average pooling is applied along the time axis:
\begin{align}
    \mathbf{E}^{\text{Speaker}} = \text{AvgPooling}(\mathbf{E}^\text{Speech}_\text{sum}) \in \mathbb{R}^{D}. \label{equ:speaker_emb}
\end{align}

The speaker verification loss consists of two components: (1) a Cross Entropy (CE) loss $\mathcal{L}_{\text{CE}}$ for speaker identification and (2) a regularization term $\mathcal{L}_{\text{Reg}}$ that bounds $\|\mathbf{w}\|$:
\begin{equation}
\resizebox{0.85\columnwidth}{!}{
$\mathcal{L}_{\text{Speaker}} = \mathcal{L}_{\text{CE}} \left( \text{Linear}(\mathbf{E}^{\text{Speaker}}), \mathbf{y}^{\text{Spk}} \right) + \beta \underbrace{\left( \| \mathbf{w} \| - 1 \right)^2}_{\mathcal{L}_{\text{Reg}}},$}
\end{equation}
where $\mathbf{y}^{\text{Spk}}$ is the ground-truth speaker label, and $\beta=0.01$ is a regularization coefficient.

The total loss for training KCE combines the CTC loss for the target-speaker ASR task and the CE loss for the SV task:
\begin{equation}
\mathcal{L}_\text{KCE} = \mathcal{L}_{\text{CTC}} + \alpha \mathcal{L}_{\text{Speaker}},
\end{equation}
where $\alpha=0.5$ balances the ASR and SV objectives.

\subsection{Speech Extraction Backbone for DAE-TSE}
The speech extraction module estimates the target speech from the mixture, conditioned on the target speaker embedding from the cue encoder. Specifically, we adopt the powerful Band-Split RNN (BSRNN)~\cite{taslp2023-luoyi-bsrnn} as the extraction backbone. 
BSRNN operates in the time-frequency domain by estimating a complex-valued spectral mask to extract the target speech from the mixture. Given the input waveform $\mathbf{x} \in \mathbb{R}^{L}$, the short-time Fourier transform (STFT) yields:
\begin{equation}
\mathbf{X} = \text{STFT}(\mathbf{x}) \in \mathbb{C}^{F \times T},
\end{equation}
where $F$ and $T$ represent the frequency bins and time frames, respectively.

BSRNN explicitly splits the mixture spectrogram into sub-bands and performs interleaved band-level and sequence-level modeling using two types of residual RNN layers. Specifically, the input spectrogram $\mathbf{X} \in \mathbb{C}^{F \times T}$ is divided into $K$ sub-band spectrograms $\mathbf{B}_s \in \mathbb{C}^{F_s \times T}$, for $s \in \{ 1, 2, \ldots, K\}$, where $F_s$ is the width of each sub-band. Independent RNNs are applied along the time and frequency dimensions in a sequential manner to model temporal and inter-band dependencies. The network then estimates a complex-valued mask $\mathbf{M} \in \mathbb{C}^{F \times T}$, which is applied to the input spectrogram via element-wise multiplication to obtain the predicted target spectrogram:
\begin{equation}
\hat{\mathbf{Y}} = \mathbf{X} \odot \mathbf{M} \in \mathbb{C}^{F \times T},
\end{equation}
where $\hat{\mathbf{Y}}$ denotes the estimated spectrogram of the target speech. Subsequently, the inverse STFT (ISTFT) is applied to reconstruct the time-domain waveform:
\begin{equation}
\hat{\mathbf{y}} = \text{ISTFT}(\hat{\mathbf{Y}}) \in \mathbb{R}^{L},
\end{equation}
where $\hat{\mathbf{y}}$ is the estimated target signal. We utilize the negative scale-invariant signal-to-noise ratio (SI-SNR)~\cite{taslp2019-luoyi-convtasnet_sisnr} as the objective function to train the extraction model:
\begin{equation}
\mathcal{L}_{\text{SI-SNR}}(\mathbf{y}, \hat{\mathbf{y}}) = -10 \log_{10} \left( \frac{ \| \mathbf{s}_{\text{target}} \|^2 }{ \| \mathbf{e}_{\text{noise}} \|^2 } \right),
\end{equation}
where
\begin{equation}
\mathbf{s}_{\text{target}} = \frac{ \langle \hat{\mathbf{y}}, \mathbf{y} \rangle }{ \| \mathbf{y} \|^2 } \mathbf{y}, \quad \mathbf{e}_{\text{noise}} = \hat{\mathbf{y}} - \mathbf{s}_{\text{target}}.
\end{equation}

The BSRNN backbone was originally proposed for blind speech separation (BSS), where no information about the target speaker is provided. To support personalized tasks such as personalized speech enhancement~(PSE) or TSE, previous work~\cite{icassp2023-jianweiyu-pse} and the open-source WeSep toolkit~\cite{is2024-shuaiwang-wesep} extend BSRNN by inserting a fusion module, allowing the network to condition its predictions on speaker embeddings. In this work, we adopt the BSRNN-based TSE implementation from WeSep to incorporate speaker information from the cue encoder into the backbone. As mentioned before, unlike conventional methods that rely on pre-enrolled utterances, our approach derives content-based speaker representations from the keywords of interest.

\subsection{DAE Paradigm: Detect, Attend and Extract}
The proposed DAE-TSE follows a three-stage detect-attend-extract formulation: the KCE first \textit{detects} the presence of keywords; if confirmed, it \textit{attends} to the target speaker; finally, the TSE backbone \textit{extracts} the target speaker's speech.
\begin{enumerate}
    \item \textbf{Detect.} DAE-TSE first detects the presence of keywords and pinpoints their temporal span through a lightweight search of the mixture. If the keyword is absent in the detection stage, the system outputs silence; otherwise, it proceeds. The cross-attention mechanism between speech and transcriptions naturally establishes frame-level correspondences between the acoustic sequence and the text-based keywords, effectively transforming detection and localization into a search problem. Leveraging this property, we develop a lightweight dynamic programming algorithm that traverses the attention matrix to efficiently detect keyword presence and localize their temporal positions in the mixture.
    Algorithm~\ref{algo:keyword-search} processes the phoneme-level cross-attention map $\mathbf{M}_{N} \in \mathbb{R}^{L_\text{kw} \times T}$ from the final KCE layer (Equation~\ref{eq:attention_map}) to compute the maximal path score $S$, the start and trigger frame indices $(i,j)$ , and a detection flag $d$. The flag $d$ is determined by thresholding $S$ against a predefined threshold $\tau$. With $O(L_\text{kw} T)$ complexity, the procedure efficiently handles typical keyword lengths.
   \item \textbf{Attend.} If keyword presence is confirmed, the DAE-TSE encoder extracts a fixed-dimension speaker embedding $\mathbf{E}^{\text{Speaker}}$ in Equation~\ref{equ:speaker_emb} through speech-text cross-attention and pooling, as described in Section~\ref{sec:kce_module}.
   \item \textbf{Extract.} With the speaker embedding $\mathbf{E}^{\text{Speaker}}$ obtained, the TSE backbone extracts the speech of the target speaker from the mixture.
\end{enumerate}

\begin{algorithm}[t!]
\caption{Detection and Localization of Keywords}
\label{algo:keyword-search}
\begin{algorithmic}[1]
\STATE \textbf{Input:} Cross-attention map $\mathbf{M}_{N}\in\mathbb{R}^{L_{\text{kw}}\times T}$, threshold $\tau$%
\STATE \textbf{Output:} Max path score $S$, start frame $i$, trigger frame $j$, detection flag $d$
\STATE $(K,T)\gets \mathbf{M}_{N}\text{.shape}$, $\text{dp}\gets\mathbf{0}^{L_{\text{kw}}\times T}$, $\text{prev}\gets\mathbf{0}^{L_{\text{kw}}\times T \times 2}$%
\FOR{$t=0,\cdots,T\!-\!1$}
    \STATE $\text{dp}[0,t]\gets\mathbf{M}_{N}[0,t]$, $\text{prev}[0,t]\gets(0,t)$
\ENDFOR
\FOR{$k=1,\cdots,K\!-\!1$}
    \FOR{$t=1,\cdots,T\!-\!1$}
        \IF{$\text{dp}[k\!-\!1,t\!-\!1]>\text{dp}[k,t\!-\!1]$}
            \STATE $\text{dp}[k,t]\gets\text{dp}[k\!-\!1,t\!-\!1]+\mathbf{M}_{N}[k,t]$
            \STATE $\text{prev}[k,t]\gets(k\!-\!1,t\!-\!1)$
        \ELSE
            \STATE $\text{dp}[k,t]\gets\text{dp}[k,t\!-\!1]+\mathbf{M}_{N}[k,t]$
            \STATE $\text{prev}[k,t]\gets(k,t\!-\!1)$
        \ENDIF
    \ENDFOR
\ENDFOR
\STATE $S\gets \max_t {\text{dp}[K\!-\!1,t]}$
\STATE $t\gets\arg\max_t\text{dp}[K\!-\!1,t]$, $k\gets K\!-\!1$
\WHILE{$k=K\!-\!1$} \STATE $(k,t)\gets\text{prev}[k,t]$ \ENDWHILE
\STATE $j\gets t\!+\!1$
\WHILE{$k>0\land t>0$} \STATE $(k,t)\gets\text{prev}[k,t]$ \ENDWHILE
\STATE $i\gets t$, $d\gets(S \geq \tau)$
\STATE \textbf{return} $S,(i,j),d$
\end{algorithmic}
\end{algorithm}

\section{Experimental Setups}

DAE-TSE is trained in two successive stages: the KCE is first optimized via the ASR-SV objective, after which the extraction backbone is trained while the KCE remains frozen.

\subsection{Data Preparation for KCE Module} 

\subsubsection{Data Simulation}  LibriSpeech~\cite{LibriSpeech} is a publicly available English speech corpus consisting of 960 hours of transcribed audio and corresponding speaker labels. For cue encoder pre-training, we use simulated mixtures generated from the train-clean-360 and train-other-500 subsets, covering 2,087 speakers. The train-clean-100 subset is excluded to prevent information leakage, as it is used for mixture generation in the backbone training.
To ensure data diversity, we adopt an online mixture generation strategy. Two clean utterances, $\mathbf{s}_1$ and $\mathbf{s}_2$, are randomly sampled and combined as:
\begin{align}
\mathbf{s}_{\text{mix}} = \gamma_1 \mathbf{s}_1 + \gamma_2 \mathbf{s}_2,
\end{align}
where the scaling factors $\gamma_1, \gamma_2 \in [0.1, 0.9]$ are randomly sampled. Online simulation adopts the LibriMix~\cite{arxiv-cosentino-librimix} max protocol: the shorter utterance is zero-padded to match the length of the longer one.

\subsubsection{Keyword Selection}
During training, we randomly crop 2-6 consecutive words from each target speaker’s transcription to form the keyword cue, comprising only 6.5\% to 19.5\% of the full transcript ($\approx$30.1 words).
During evaluation, we restrict the reference to at most 4 successive words—covering $\le28\%$ of the 14.1-word average transcript.
The selected words are converted to phoneme sequences with a grapheme-to-phoneme toolkit \cite{arxiv-lee-g2p} and fed to the cue encoder.

\subsubsection{Evaluation Dataset Simulation}

To comprehensively evaluate the keyword attending and detection capabilities, we consider both scenarios: when the keyword is present and absent in the mixture. Specifically, we construct an evaluation set consisting of 2,620 test samples, each paired with a randomly sampled fixed keyword cue. Approximately 50\% of the mixtures contain keywords that do not correspond to any speaker in the mixture, simulating cases where the queried keyword is absent from the speech.

\subsection{Data Preparation for Extraction Backbone}
To evaluate the performance of the DAE-TSE backbone, we conduct experiments on the Libri2Mix dataset~\cite{arxiv-cosentino-librimix}, a two-speaker mixture corpus derived from LibriSpeech. The training set is simulated using the train-clean-100 subset, while the validation and test sets are generated from the dev-clean and test-clean subsets, respectively. Specifically, we adopt the fully overlapped (min) version sampled at 16 kHz.

As mentioned before, the speakers used for cue encoder pre-training are disjoint from those used in backbone training and evaluation, ensuring a fair assessment of the generalization and effectiveness of the proposed DAE-TSE system.

\begin{table*}[ht]

    \centering
    \begin{resizebox}{2.0\columnwidth}{!}{
    \begin{tabular}{ll|ccccccccc}
        \toprule
        \multicolumn{1}{c}{\textbf{Model}} & \textbf{Cue Info.} & \textbf{SI-SNRi (dB)$\uparrow$} & \textbf{Acc. (\%)$\uparrow$} & \textbf{PESQ$\uparrow$} & \textbf{STOI$\uparrow$} & \textbf{DNSMOS$\uparrow$} & \textbf{SPK\_SIM$\uparrow$} & \textbf{dWER (\%)$\downarrow$} \\
        \midrule
            \multirow{2}{*}{TSE} & Pipeline & 12.98 & 89.93 & 2.84 & 88.72 & \textbf{3.17} & 0.964 & 20.81 \\
            & Audio & 13.52 & 90.83 & 2.86 & 89.51 & 3.15 & 0.967 & 19.84 \\
            Multi-Level TSE~\cite{icassp2025-zhangke-multi_level_speaker_emb} & Audio & 16.08 & 97.18 & 2.86 & 93.71 & 3.15 & 0.978 & 15.14 \\
        \midrule
            DAE-TSE & Keywords (k=4) & \textbf{16.45} & \textbf{98.98} & \textbf{2.87} & \textbf{95.18} & 3.14 & \textbf{0.982} & \textbf{13.62} \\
        \bottomrule
    \end{tabular}
    }\end{resizebox}
    \caption{Comparison of DAE-TSE with representative TSE baselines on Libri2Mix. DAE-TSE uses consecutive keywords, which are randomly sampled from the transcription, as the enrollment cue. The \textit{Audio} setup uses a clean enrollment utterance from the target speaker, while the \textit{Pipeline} baseline refers to a cascaded system of a speech separation front-end and ASR backend. Best results are in \textbf{bold}.}
    \label{tab:main}
\end{table*}

\subsection{Training Details}
The KCE module is first trained for 150 epochs (10-epoch warm-up) with Adam~\cite{iclr2015-adam-Diederik} at a learning rate of 1e-3 and a batch size of 64. Input features are 80-dimensional FBank coefficients (25 ms window, 10 ms shift). For the TSE backbone, we adopt the BSRNN implemented in the WeSep framework~\cite{is2024-shuaiwang-wesep}, as it achieves the best overall performance. Full-length mixtures are fed to the BSRNN extractor without chunking. STFT uses a 16 kHz sampling rate, a 512-sample window, and a 128-sample hop, yielding 128 frequency bins. Subsequently, the TSE backbone is trained for 150 epochs while the cue encoder remains frozen; the learning rate decays exponentially from \texttt{1e-3} to \texttt{2.5e-5} without warm-up. All experiments run on 8 NVIDIA V100 GPUs.

\subsection{Baselines}
\label{sec:baseline}
Following the open-source setup in~\cite{is2024-shuaiwang-wesep} for the BSRNN-based TSE system, all extraction-based baselines adopt the same ECAPA-TDNN~\cite{is20-desplanques20-ecapa_tdnn} model pretrained on VoxCeleb2~\cite{is2018-chung-voxceleb2} in the WeSpeaker toolkit~\cite{icassp2023-wanghongji-wespeaker}, which is publicly available\footnote{\url{https://wenet.org.cn/downloads?models=wespeaker\&version=voxceleb\_ECAPA512.zip}}.
Moreover, since the proposed DAE-TSE leverages keywords as cues, we compare it not only with standard TSE systems but also with a cascaded pipeline that sequentially applies BSS and ASR.
Specifically, from the perspective of the type of information utilized, such as speaker identity or content-based keywords, we consider the following baselines.
\begin{itemize}
    \item \textbf{Audio (Standard)}. The standard TSE system utilizes a clean enrollment utterance from the target speaker to extract a speaker embedding, which serves as a conditioning vector for extracting the target speech from a multi-speaker mixture. This method requires clean enrollment data and relies solely on speaker identity, without leveraging any content-related information. It represents the most widely adopted baseline in TSE research.

    \item \textbf{Pipeline}.
    The pipeline baseline is a cascaded system consisting of a separation front-end and an ASR backend: the mixture is first decomposed into multiple waveforms, each of which is transcribed; the channel with the minimum edit distance to the enrolled keywords is selected, and its separated speech is adopted as the clean enrollment utterance.
    Specifically, we adopt the powerful open-source MossFormer~\cite{icassp2024-zhaoshengkui-mossformer2}, which achieves state-of-the-art performance on Libri2Mix. Separated sources are transcribed by a high-performance CTC-Transducer-based ASR model\footnote{\url{https://catalog.ngc.nvidia.com/orgs/nvidia/teams/nemo/models/stt\_en\_fastconformer\_hybrid\_large\_pc}} from NVIDIA NeMo~\cite{nemo} Team.
\end{itemize}

\subsection{Evaluation Metrics}
To comprehensively evaluate the performance of all TSE systems, we adopt a wide range of evaluation metrics from different aspects of evaluation:
\begin{itemize}
    \item \textbf{PESQ and STOI}: Perceptual Evaluation of Speech Quality (PESQ) and Short-Time Objective Intelligibility (STOI) are full-reference metrics that estimate perceived speech quality and speech intelligibility under noisy conditions, respectively.
    
    \item \textbf{DNSMOS}: DNSMOS is a reference-free perceptual quality estimator designed for 16 kHz audio, which produces three scores in the range of 1 to 5: SIG (signal quality), BAK (background noise quality), and OVRL (overall quality). For simplicity, we report only the OVRL score.
    
    \textbf{SI-SNR improvement (SI-SNRi)}: SI-SNRi measures enhancement in speech quality relative to the mixture.
    
    \item \textbf{Accuracy}: Enhancement accuracy~\cite{icassp2025-zhangke-multi_level_speaker_emb} is defined as the percentage of trials whose SI-SNRi exceeds 1 dB, a threshold that designates successful target-speaker extraction.
    
    \item \textbf{SPK-SIM}: Speaker similarity is computed as the cosine similarity between speaker embeddings extracted from the enhanced and reference signals. We use a pre-trained WavLM model~\cite{jstlp2022-sanyuanchen-sanyuan} for evaluation.
    
    \item \textbf{Differential Word Error Rate (dWER)}: This metric measures the word error rate (WER) between the ASR transcription of the extracted speech and that of the ground-truth target speech. We employ the base version of Whisper for ASR decoding.
\end{itemize}

Additionally, for the keyword detection and localization evaluation of KCE, samples containing the keyword are treated as positive examples, while those without the keyword are considered negative examples. Detection performance is reported via precision (Pre.), recall (Rec.), and F1-score; localization accuracy is measured by the $\ell_1$ error of the start (S Err.) and end (E Err.) timestamps.

\section{Results and Analysis}
\subsection{Performance of Keyword Guided TSE}

Table~\ref{tab:main} presents the extraction results for several TSE baselines and various DAE-TSE configurations.
Compared to the standard baselines, our DAE-TSE achieves superior performance with only 4 keywords (approximately 28.4\% of the full transcription) across nearly all evaluation metrics when compared to our baseline and the powerful Multi-layer TSE. DAE-TSE demonstrates strong performance not only in signal quality metrics (SI-SNRi, PESQ, STOI) but also in acoustic and semantic consistency (Accuracy, SPK-SIM, dWER). These results highlight the effectiveness of DAE-TSE in leveraging local linguistic cues to infer global speaker and contextual information, thereby enhancing the overall extraction capability of the TSE system.

\subsection{Keyword Presence Detection and Temporal Localization}
DAE-TSE considers a common yet realistic scenario in which the provided keywords may not appear in the target speech. 
\begin{table}[hbt]
    \centering
    \begin{tabular}{@{}S[table-format=1.2]*{3}{S[table-format=2.2]}S[table-format=3.1]S[table-format=3.1]@{}}
        \toprule
        $\tau$ & \multicolumn{3}{c}{\textbf{Detection Metrics (\%)}} & \multicolumn{2}{c}{\textbf{Time Error (ms)}} \\
        \cmidrule(lr){2-4} \cmidrule(lr){5-6}
        & {\textbf{Pre.}} & {\textbf{Rec.}} & {\textbf{F1}} & {\textbf{S Err.}} & {\textbf{E Err.}} \\
        \midrule
        0.48 & \textbf{99.00} & 93.85 & 96.36 & 103.7 & 100.6 \\
        0.36 & 98.64 & 97.24 & 97.93 & 104.0 & 100.7 \\
        0.33 & 98.49 & 97.63 & \textbf{98.06} & \textbf{103.7} & 100.4 \\
        0.30 & 97.79 & 97.71 & 97.75 & 103.6 & \textbf{100.3} \\
        0.23 & 96.88 & \textbf{98.11} & 97.49 & 104.3 & 101.0 \\
        \bottomrule
    \end{tabular}
    \caption{Keyword presence detection and localization results under different thresholds. $\tau$ denotes the threshold. Pre., Rec., and F1 denote Precision, Recall, and F1-score, respectively. S Err. and E Err. represent the absolute time shift error of the start and end point when the keyword is present in the mixture. Best results are in \textbf{bold}.}
    \label{tab:presence-and-localization}
\end{table}
Consequently, DAE-TSE first decides whether the keywords are present in the mixture. If detection is positive, DAE-TSE further attends to and extracts the target speaker; otherwise, it simply outputs silence. When multiple speakers utter the keyword, DAE-TSE randomly selects one as the target.
As shown in Table~\ref{tab:presence-and-localization}, the proposed detection algorithm effectively identifies keyword presence and achieves strong detection performance. Additionally, the localization errors are minimal, demonstrating that the model can accurately determine the temporal positions of the keywords. This capability significantly enhances the practicality and versatility of the proposed DAE-TSE framework.

\begin{figure}[ht]
  \centering
  \includegraphics[width=1.0\linewidth]{figures/attention_map_visualization.png}
  \caption{Cross-attention heatmaps for a positive sample (left, keywords present) and a negative sample (right, keywords absent). The horizontal axis denotes speech frame indices, while the vertical axis represents the keyword phoneme sequence.}
  \label{fig:attention_map}
\end{figure}

We further visualize the cross-attention map from the last layer of KCE in Figure~\ref{fig:attention_map}. When the keyword is present in the mixture, the attention map exhibits a clear continuous highlight, indicating a strong alignment pattern between the mixture frames and the keyword sequence.
In contrast, when the keyword is absent, the attention map appears scattered without a consistent alignment. These observations demonstrate that the cue encoder effectively captures the correlation between the mixture and the provided keywords, thereby bridging keyword-to-speaker alignment. This property enables our framework to perform keyword presence detection and temporal localization. Additional attention heatmaps are provided on the demo page.

\subsection{Ablation Studies}
\subsubsection{Training Objectives for Cue Encoder}
Table~\ref{tab:encoder_loss} quantifies the impact of each loss of KCE pretraining on the TSE performance. Omitting the regularization term $\mathcal{L}_{\text{Reg}}$ from the speaker loss $\mathcal{L}_{\text{Speaker}}$ incurs minor degradation; completely removing $\mathcal{L}_{\text{Speaker}}$ produces a marked drop in SI-SNRi. These ablations confirm that speaker-oriented losses are indispensable for compelling the encoder to capture reliable speaker signatures. Moreover, removing the ASR loss $\mathcal{L}_{\text{CTC}}$ from keyword-guided cue encoder training precipitates a pronounced performance drop, evidencing that the original target-speaker embedding encodes indispensable keyword context. This semantic content is as vital as the identity itself for accurate extraction. Collectively, the ablations underscore the core strength of KCE: a speaker encoder that captures the context of keywords through alignment between mixture speech and keyword enrollment.

\begin{table}[ht]
    \centering
    \begin{tabular}{l|cc}
    \toprule
    \textbf{Model} & \textbf{SI-SNRi (dB)} & \textbf{Acc. (\%)} \\
    \midrule
        DAE-TSE & \textbf{16.45} & \textbf{98.98} \\
        {~~~~}w/o $\mathcal{L}_{\text{Reg}}$ & 15.97 & 98.02 \\
        {~~~~}w/o $\mathcal{L}_{\text{Speaker}}$ & 14.72 & 96.13 \\
        {~~~~}w/o $\mathcal{L}_{\text{CTC}}$ & 3.47 & 70.15 \\
        \bottomrule
    \end{tabular}
    \caption{Ablation results of removing different components from the training objective of the cue encoder. Specifically, $\mathcal{L}_{\text{Reg}}$ denotes the regularization term in the speaker verification loss, $\mathcal{L}_{\text{Speaker}}$ represents the speaker prediction loss, and $\mathcal{L}_{\text{CTC}}$ corresponds to the keyword-aware ASR objective.}
    \label{tab:encoder_loss}
\end{table}

\subsubsection{Keyword Lengths for Cue Encoder}
As shown in Table~\ref{tab:ablation-kw-length}, we investigate how the number of keywords used as cues influences separation performance.
The results show a clear monotonic improvement as the number of keywords increases. Even with just a single keyword, the performance remains strong and surpasses that of the standard TSE system reported in Table~\ref{tab:main}. Longer keyword sequences provide richer contextual information, further improving extraction quality. Notably, when using approximately 4 keywords, the performance gap compared to using the full transcription becomes negligible. These consistent ablation results confirm that the speaker embedding captures rich global contextual information and that KCE is capable of extracting this information effectively using only a few keywords.

\begin{table}[h]
    \centering
    \begin{tabular}{c|cc}
        \toprule
        \textbf{\#Keywords} & \textbf{SI-SNRi (dB)} & \textbf{Acc. (\%)} \\
        \midrule
            1 & 15.34 & 96.33 \\
            2 & 15.74 & 97.18 \\
            3 & 16.37 & 98.70 \\
            4 & 16.45 & \textbf{98.98} \\
        \midrule
            Full Trans. & \textbf{16.48} & 98.78 \\
        \bottomrule
    \end{tabular}
    \caption{Results for different keyword lengths (\#Keywords) used as enrollment cues. Additionally, we evaluate the DAE-TSE system with full transcription enrollment (Full Trans.), where each transcription contains an average of 14.1 words.}
    \label{tab:ablation-kw-length}
\end{table}

\subsection{Visualization of Cue Embeddings}

\begin{figure}[ht]
  \centering
  \includegraphics[width=1.0\linewidth]{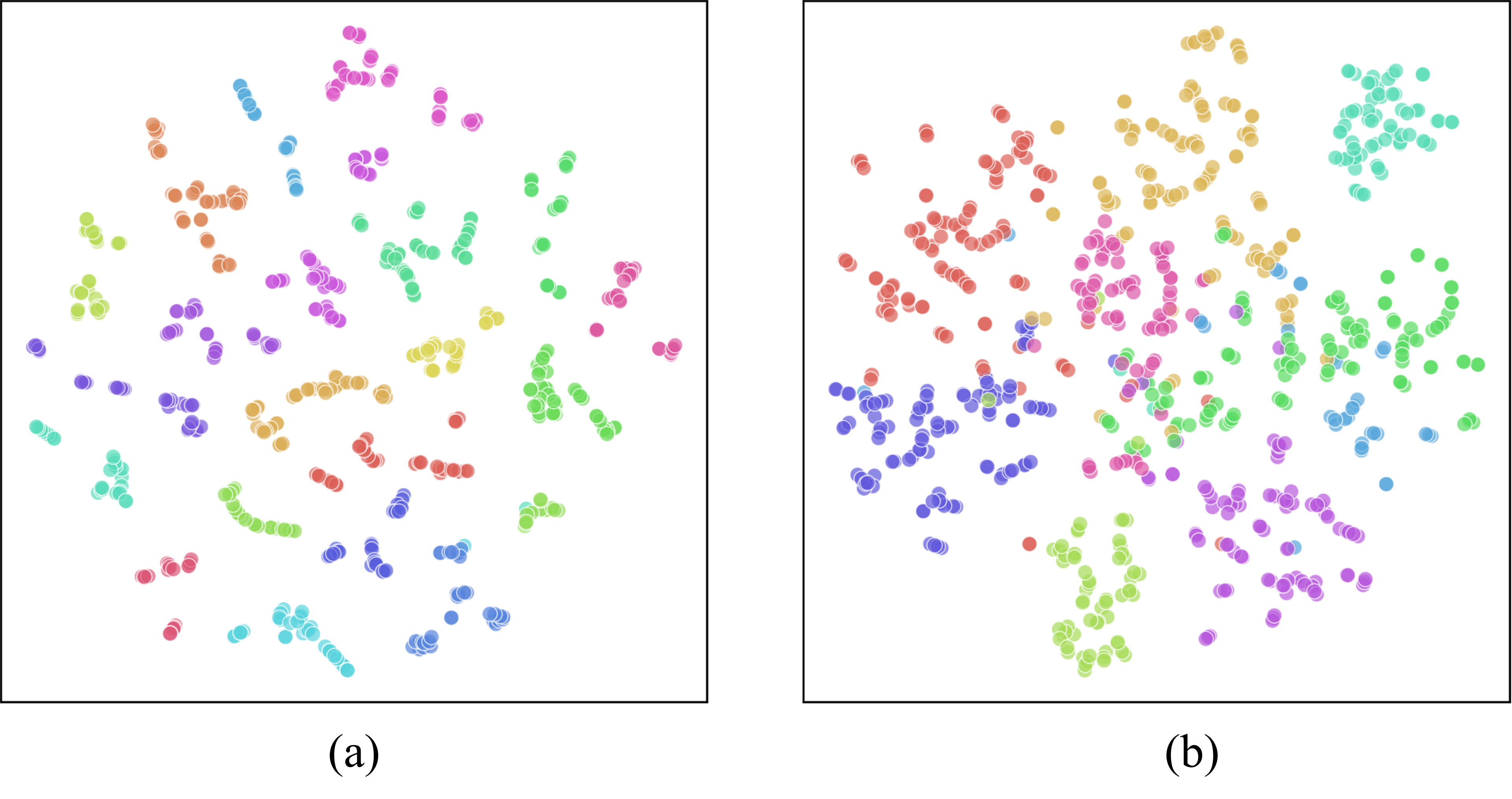}
  \caption{t-SNE scatter plot of speaker embeddings, with each color representing a target speaker. (a) Embeddings in the same color are extracted from the same mixture across different keywords. (b) Embeddings in the same color are extracted from different mixtures.}
  \label{fig:tsne}
\end{figure}

In this section, we visualize the distribution of cue embeddings using t-SNE~\cite{maaten2008visualizing}. We analyze two scenarios: in Figure~\ref{fig:tsne}(a), points with the same color originate from the same mixture using different keyword cues; in Figure~\ref{fig:tsne}(b), points with the same color are extracted from different mixtures.

In both cases, each cluster generally aligns with a single speaker, indicating the speaker-discriminative nature of the cue embeddings. The key distinction lies in contextual consistency: in (a), the embeddings are drawn from the same mixture and thus share a consistent acoustic and linguistic context, whereas in (b), the embeddings lack shared context because they are derived from different mixtures.

In Figure~\ref{fig:tsne}(a), we observe well-separated clusters with large inter-cluster distances and tight intra-cluster cohesion. This suggests that both the identity of the speaker and the consistent contextual information contribute significantly to the quality of the cue embeddings and the effectiveness of the representation of the target speaker. In contrast, Figure~\ref{fig:tsne}(b) shows that even in the absence of shared context, where embeddings are extracted from different mixtures, clusters still largely align with speaker identity. However, some overlap near cluster boundaries indicates that the lack of contextual consistency may introduce ambiguity. These results demonstrate that local keyword-based cues can effectively capture global contextual information, thereby supporting robust target speaker extraction.

\section{Conclusion}
In this work, we propose DAE-TSE, a novel keyword-guided target speaker extraction framework that leverages partial transcriptions as reference cues.
DAE-TSE establishes a three-stage paradigm for keyword-guided TSE: detecting the presence of keywords, attending to the target speaker for representation, and extracting the target speech. The design of DAE-TSE not only enables target speaker extraction but also provides accurate keyword presence detection and temporal localization capabilities via the cross-attention interaction between the mixture and the keyword cues.
Extensive experiments on Libri2Mix demonstrate that DAE-TSE outperforms standard TSE baselines that rely on pre-enrolled speech. To the best of our knowledge, this work is the first to explore the use of partial transcriptions to specify the target speaker in TSE, paving the way for more flexible and real-world speech extraction systems.
To advance the framework toward practical application, future efforts will focus on evaluating its robustness in challenging environments, such as scenarios with more speakers, sparse overlap, and recordings with background noise and reverberation.

\newpage

\section*{Acknowledgements}

This work was supported by the National Natural Science Foundation of China (Grant No. 62401377) and the Yangtze River Delta Science and Technology Innovation Community Joint Research Project (Grant No. 2024CSJGG1100).
This work was also supported by the National Natural Science Foundation of China (Grant No. 92370206).

\section*{Contribution Statement}

Haoyu Li$^{*}$ and Yu Xi$^{*}$ contributed equally to this work.

\bibliographystyle{named}
\bibliography{refs}

\end{document}